\documentclass[doublecol,figures]{epl2}
\usepackage{amssymb}
\begin{document}

\title{Particle displacements in the {elastic} deformation of amorphous materials:
  local fluctuations vs.\ non-affine field}
\shorttitle{Particle displacements in the {elastic} deformation of amorphous materials}
\author{C. Goldenberg\inst{1}\thanks{E-mail: \email{chayg@pmmh.espci.fr}} \and 
A. Tanguy\inst{2}\thanks{E-mail: \email{atanguy@lpmcn.univ-lyon1.fr}} \and 
J.-L. Barrat\inst{2}\thanks{E-mail: \email{barrat@lpmcn.univ-lyon1.fr}}}
\shortauthor{C. Goldenberg \etal}
\institute{
\inst{1} Laboratoire de Physique et M\'ecanique des Milieux
  H\'et\'erog\`enes (CNRS UMR 7636), ESPCI\\ 10 rue Vauquelin, F-75231 Paris Cedex
  05, France.\\
\inst{2} Laboratoire de Physique de la Mati{\`e}re Condens{\'e}e et Nanostructures,
Univ. Lyon; Univ. Lyon 1; CNRS UMR 5586\\ Domaine Scientifique de la Doua, F-69622
Villeurbanne Cedex, France.
}

\abstract{%
  We study the local disorder in the deformation of amorphous materials by
  decomposing the particle displacements into a continuous, inhomogeneous field
  and the corresponding fluctuations.  We compare these fields to the commonly
  used non-affine displacements, in an elastically deformed 2D Lennard-Jones
  glass. Unlike the non-affine field, the fluctuations are very localized, and
  exhibit a much smaller (and system size independent) correlation length,
  {on the order of a} particle diameter, {supporting the
    applicability of the notion of local ``defects'' to such materials}. We
  propose a scalar ``noise'' field to characterize the fluctuations, as an
  additional field for extended continuum models, e.g., to describe the
  localized irreversible events observed during plastic deformation.}


\maketitle

\section{Introduction}
The nature of fluctuations in glasses and other amorphous materials out of
equilibrium is of much current interest. While elasticity and plasticity are
often employed for describing both crystalline and amorphous materials, their
microscopic basis is well-established only in crystalline (or polycrystalline)
materials, and relies on the periodicity of the microscopic structure (possibly
with localized defects)~\cite{Born88,Cottrell53}. As in crystal plasticity,
localized rearrangements appear to play an important role in the plastic
deformation of amorphous materials~\cite{Falk98,Maloney06,Tanguy06}, but the
lack of underlying order renders the identifications of localized ``defects''
difficult. In crystals (with a simple unit cell) under homogeneous deformation,
the particle displacements conform to the imposed (affine) strain, but in
amorphous materials they do not~\cite{Alexander98}. The {\em non-affine}
displacements (obtained by subtracting the expected homogeneous deformation)
have recently been studied in experiments and simulations of different
amorphous systems (e.g., glasses, colloids, granular materials and
foams~\cite{Langer97,Tanguy02,Radjai02,%
  Leonforte04,Leonforte05,DiDonna05,Maloney06b,Ellenbroek06}).  They are
typically of the same order of magnitude as the relative affine displacements
of neighboring particles, and therefore cannot be considered a small
correction: ignoring them, or treating them as a perturbation, yields highly
inaccurate estimates for macroscopic material properties such as the elastic
moduli~\cite{Schwartz84,Liao97,Tanguy02}. Considering the non-affine
displacements as a fluctuation, or ``noise''~\cite{Tanguy02,Leonforte04}, poses
difficulties since they exhibit long range
correlations~\cite{Tanguy02,Leonforte04,Leonforte05,DiDonna05,Leonforte06,Maloney06b}
which would render the contribution of such ``noise'' dominant at large scale.

In this Letter we show, using numerical simulations of a two dimensional (2D)
Lennard-Jones glass subject to small elastic deformation, that {the
  main features} of the non-affine field can be captured by a continuous,
inhomogeneous (subsystem scale) displacement field. However, the microscopic
displacements exhibit significant fluctuations with respect to this field. A
full characterization of the displacements therefore requires not only a
distinction between an affine and non-affine contribution, but also between a
continuous and a fluctuating part. We present the first study of the local
fluctuations, and show that their properties are very different from those of
the non-affine field: they are essentially uncorrelated and extremely
localized; their distribution is qualitatively different. Furthermore, unlike
the non-affine displacement, whose correlation depends on the system size
(see~\cite{Maloney06b} and below), the (very short) correlation of the
fluctuations does not. We propose a new continuum scalar ``noise'' field based
on the fluctuations. This field is formally analogous to the kinetic
temperature in the kinetic theory of gases, which quantifies the local
deviations from the coarse grained (hydrodynamic) velocity field. The ``noise''
field may be used for supplementing the macroscopic displacement field in
extended continuum descriptions, e.g., to account for the part of the
microscopic elastic energy not captured by the macroscopic
field~\cite{Goldhirsch02,Serero07}.  The localized nature of the fluctuations,
and patterns observed in the ``noise'' field parallel to the principal shear
directions, suggest a relation (confirmed by preliminary
results~\cite{GoldenbergTanguyBarratIP}) between the fluctuations in the
elastic response and irreversible localized rearrangements observed in plastic
deformation~\cite{Maloney06,Tanguy06}.

\section{Definitions}
Consider particles (labeled by Roman indices) whose center of mass positions
are denoted $\left\{\vect{r}_{i}^{0}\right\}$ in a reference configuration and
$\left\{\vect{r}_{i}\right\}$ in a deformed one.  The displacement of particle
$i$ is $\vect{u}_{i}\equiv \vect{r}_{i} - \vect{r}_{i}^{0}$. In a simple
periodic lattice (with one particle per unit cell), under a uniform strain
$\tens{\epsilon}$, the deformation is {\em affine}, i.e., the relative displacement of
two particles, $\vect{u}_{ij}=\vect{u}_{i}-\vect{u}_{j}$, is given by:
\begin{equation}
  u_{ij\alpha}=\epsilon_{\alpha\beta}r_{ij\beta}^{0}\,,
  \label{eq:affine-displacement}
\end{equation}
where $\vect{r}_{ij}^{0}=\vect{r}_{i}^{0}-\vect{r}_{j}^{0}$ and Greek indices
denote Cartesian coordinates (the Einstein summation convention is used).  This
relation between the particle displacements and the macroscopic deformation
forms the microscopic basis of crystal elasticity~\cite{Born88}.  It has long
been appreciated~\cite{Alexander98} that in disordered systems, the
displacements are not affine even under uniform applied strain, since an affine
deformation would result in a configuration in which force balance is violated.
The non-affine particle displacements,
\begin{equation}
  \delta u_{i\alpha}\equiv u_{i\alpha}-\epsilon_{\alpha\beta}r_{i\beta}^{0}\,,
  \label{eq:nonaffine-displacement}
\end{equation}
typically exhibit long range
correlations~\cite{Tanguy02,Leonforte04,Leonforte05,Leonforte06,Maloney06b},
which depend on the system size (as observed in~\cite{Maloney06b} and discussed
further below): the correlation of {their scalar product} crosses zero
at a distance of about 0.3 times the system width.  {This can be
  explained by treating the material as inhomogeneously elastic, which predicts
  a power law decay of the correlations (logarithmic in
  2D)~\cite{DiDonna05,Maloney06b}}.  {Therefore, rather than treating
  the non-affine displacements as discrete, particle scale fluctuations, their
  main features may be captured by an inhomogeneous, subsystem scale continuum
  description.}

{In order to define microscopic displacement fluctuations, one needs
  to subtract a local, continuous displacement field from the individual
  particle displacements.  Rather than performing an arbitrary smoothing or
  interpolation of the particle displacements,} we employ a definition proposed
in~\cite{Goldhirsch02}, based on a {systematic} spatial coarse
graining (CG) procedure. This procedure uses the standard definition of the
microscopic densities of mass, momentum and energy as sums over Dirac delta
functions centered at the particles' centers of mass. The corresponding CG
densities are obtained by a convolution with a spatial CG function
$\phi(\vect{R})$, a normalized non-negative function with a single maximum at
$\vect{R}=0$ and a characteristic width $w$, the CG scale (e.g., a Gaussian: in
2D, \mbox{$\phi(\vect{R})=\frac{1}{\pi w^2}e^{-(|\vect{R}|/w)^2}$)}.  In
particular, the CG mass density is $\rho(\vect{r},t)\equiv\sum_i m_i
\phi[\vect{r}-\vect{r}_i(t)]$, and the CG momentum density is $\vect{p}({\bf
  r},t)\equiv \sum_i m_i \vect{v}_i(t) \phi[\vect{r}-\vect{r}_i(t)]$, where
$\left\{m_{i}\right\}, \left\{\vect{v}_{i}(t)\right\}$ are the particle masses
and velocities at time $t$, respectively. The definition of the CG displacement
follows its standard definition in continuum mechanics, i.e., the Lagrangian
time integral over the velocity of a {\em material particle}. The CG velocity
is given by:
\begin{equation}
  \label{eq:velocity}
  \vect{v}(\vect{r},t)\equiv\frac{\vect{p}(\vect{r},t)}{\rho(\vect{r},t)} = 
  \frac{\sum_i m_i \vect{v}_i(t)
    \phi[\vect{r}-\vect{r}_i(t)]}{\sum_j m_j \phi[\vect{r}-\vect{r}_j(t)]}\,.
\end{equation}
Therefore, the CG displacement field is given by~\cite{Goldhirsch02,Kolb06}:
\begin{eqnarray}
  \label{eq:cg-displacement}
  \vect{U}(\vect{R},t) &\equiv&
  \int_0^t \vect{v}[\vect{r}(\vect{R},t'),t']\,\upd t'\nonumber\\
  &=&\int_0^t \frac{\sum_{i} m_i \vect{v}_i(t')
    \phi[\vect{r}(\vect{R},t')-\vect{r}_i(t')]} 
  {\sum_{j} m_j \phi[\vect{r}(\vect{R},t')-\vect{r}_j(t')]}\,\upd t'\nonumber\\
  &=&   \frac{\sum_{i} m_i \vect{u}_i(t)
    \phi[\vect{R}-\vect{r}^0_i]} 
  {\sum_{j} m_j \phi[\vect{R}-\vect{r}^0_j]}+\mathcal{O}\left(\epsilon^2\right),
\end{eqnarray}
where $\vect{R}$ denotes the Lagrangian coordinate of a material particle whose
(Eulerian) coordinate at time $t$ is $\vect{r}$, $\vect{u}_i(t)$ is the total
displacement of particle $i$, and $\epsilon$ is a measure of the local strain;
the {last line} is obtained using integration by
parts~\cite{Goldhirsch02}. {The particle masses in
  Eq.~(\ref{eq:cg-displacement})} ensure consistency with a dynamical (time
dependent) description~\cite{Goldhirsch02}; for the quasi-static deformation
considered here, $t$ may be {interpreted} as a parameter
characterizing the overall deformation. For small strain, our focus in this
Letter, one can use the approximate expression on the {last} line of
Eq.~(\ref{eq:cg-displacement}), which is trajectory independent, i.e., it
involves only the total particle displacements. {With a smooth
  $\phi(\vect{R})$, Eq.~(\ref{eq:cg-displacement}) defines a smooth
  displacement field which can be used to calculate local, scale dependent
  strain measures based on its gradients}. Unlike strain definitions based on a
local fit to an affine deformation~(e.g.,~\cite{Liao97,Falk98}), those are, by
construction, fully consistent with continuum mechanics, and may be extended to
large deformations.  The CG displacement is defined only in terms of the
particle displacements, and does not rely on any assumptions regarding material
homogeneity or constitutive description, as does the decomposition into an
affine and non-affine part.  It can therefore also be used for describing
inelastic deformation (e.g., plastic flow or fracture).  Based on
Eq.~(\ref{eq:cg-displacement}), we define the {\em displacement fluctuations}
as $\vect{u}'_i(\vect{r})\equiv \vect{u}_i - \vect{U}(\vect{r})$. For
comparison with the non-affine field, it is useful to consider the fluctuations
at the particle positions, $\vect{u}'_i(\vect{r}^0_i)$, denoted below as
$\vect{u}'_i$.  The CG displacement field, $\vect{U}(\vect{r})$, and hence the
fluctuations $\vect{u}'_i$, are in general {\em scale dependent}. A possible
definition of a corresponding CG scalar ``noise'' field is
\mbox{$\eta(\vect{r})\equiv \sum_i m_i |\vect{u}'_i(\vect{r})|^2
  \phi[\vect{r}-\vect{r}_i(t)]$}. This expression is formally similar to the
kinetic temperature in the kinetic theory of gases~\cite{Serero07} with the
velocity fluctuations replaced by the displacements fluctuations (except for a
density factor; the density is quite homogeneous in the systems considered
here).

\section{Results}
We apply the above definitions to 2D amorphous
solids~\cite{Tanguy02,Leonforte04} prepared by quenching a fluid of
polydisperse particles (with diameters $\left\{\sigma_{i}\right\}$ uniformly
distributed in the range \mbox{$0.8 \sigma - 1.2 \sigma$}) of equal (unit)
mass, interacting via a Lennard-Jones potential, \mbox{$V_{ij}(r)=4\varepsilon
  \left[(\sigma_{ij}/r)^{12}-(\sigma_{ij}/r)^6\right]$}, where
\mbox{$\sigma_{ij}=(\sigma_{i}+\sigma_{j})/2$}. The mean particle diameter
$\left< \sigma \right>$ defines the unit of length. We use periodic boundary
conditions in both directions.  We apply three different uniform deformation
{modes}: uniaxial stretching parallel to the $x$ or $y$ axis, and a
simple shear parallel to the $x$ axis.  For a uniaxial deformation, the length
of the simulation cell is multiplied by $1+\epsilon$, while for simple shear,
we use Lees-Edwards boundary conditions with shear strain $\gamma=\epsilon/2$.
We impose the affine displacement [Eq.~(\ref{eq:affine-displacement})] on each
particle, and subsequently relax the system to the nearest energy
minimum~\cite{Leonforte05}.  We use $\epsilon=10^{-6}$, which ensures a linear
response~\cite{Tanguy06}. Unless noted otherwise, the data presented were
obtained using systems with $N=10000$ particles of size $104\times 104$ (i.e.,
the mean density is $\rho=0.925$).

\begin{figure}[h!]
\raisebox{0.4\hsize}{\bf a}\hspace{0.006\hsize}
    \includegraphics[width=0.435\hsize,clip]{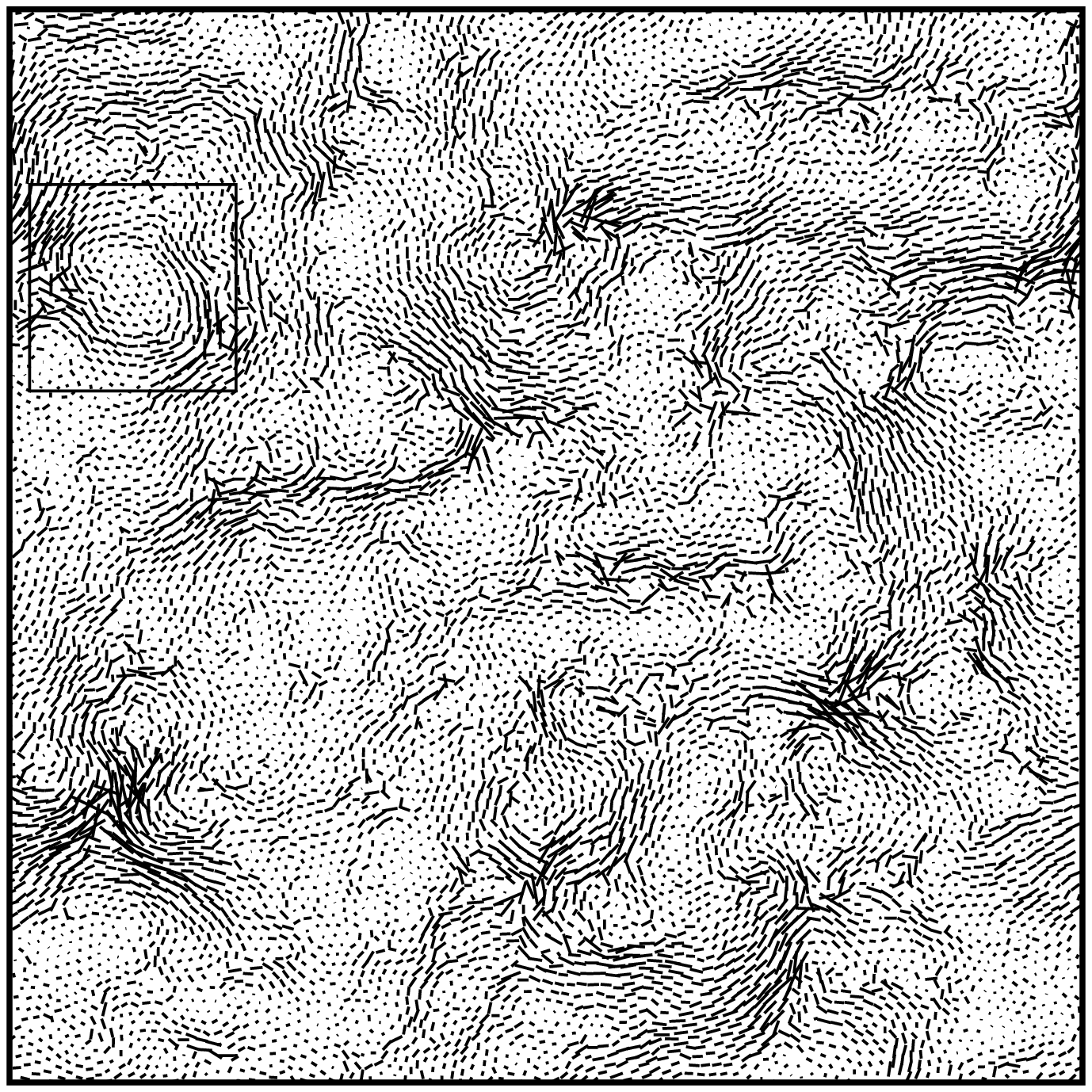}
\raisebox{0.4\hsize}{\bf b}\hspace{0.017\hsize}
    \includegraphics[width=0.435\hsize,clip]{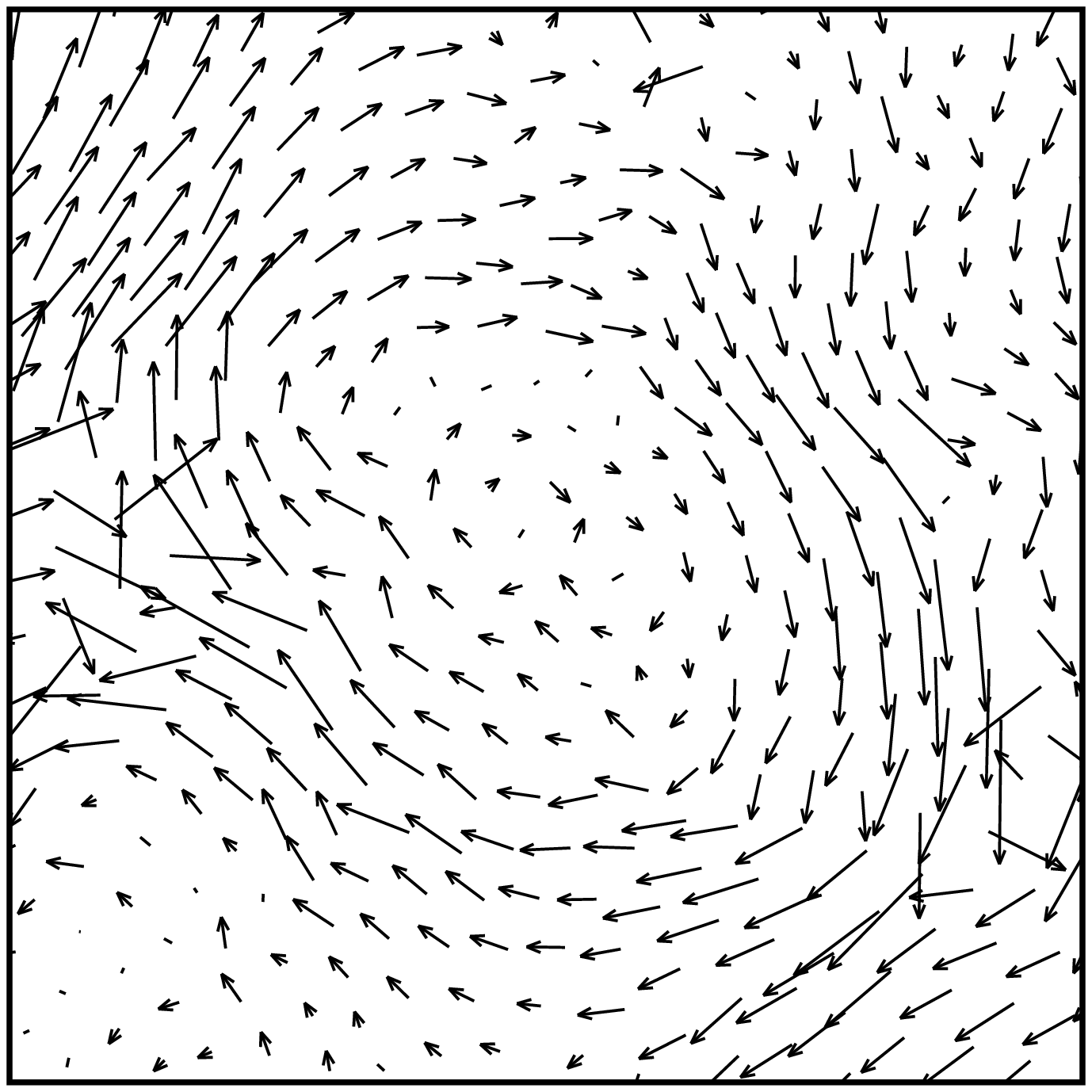}\\
\raisebox{0.435\hsize}{\bf c}
  \includegraphics[width=0.45\hsize,clip]{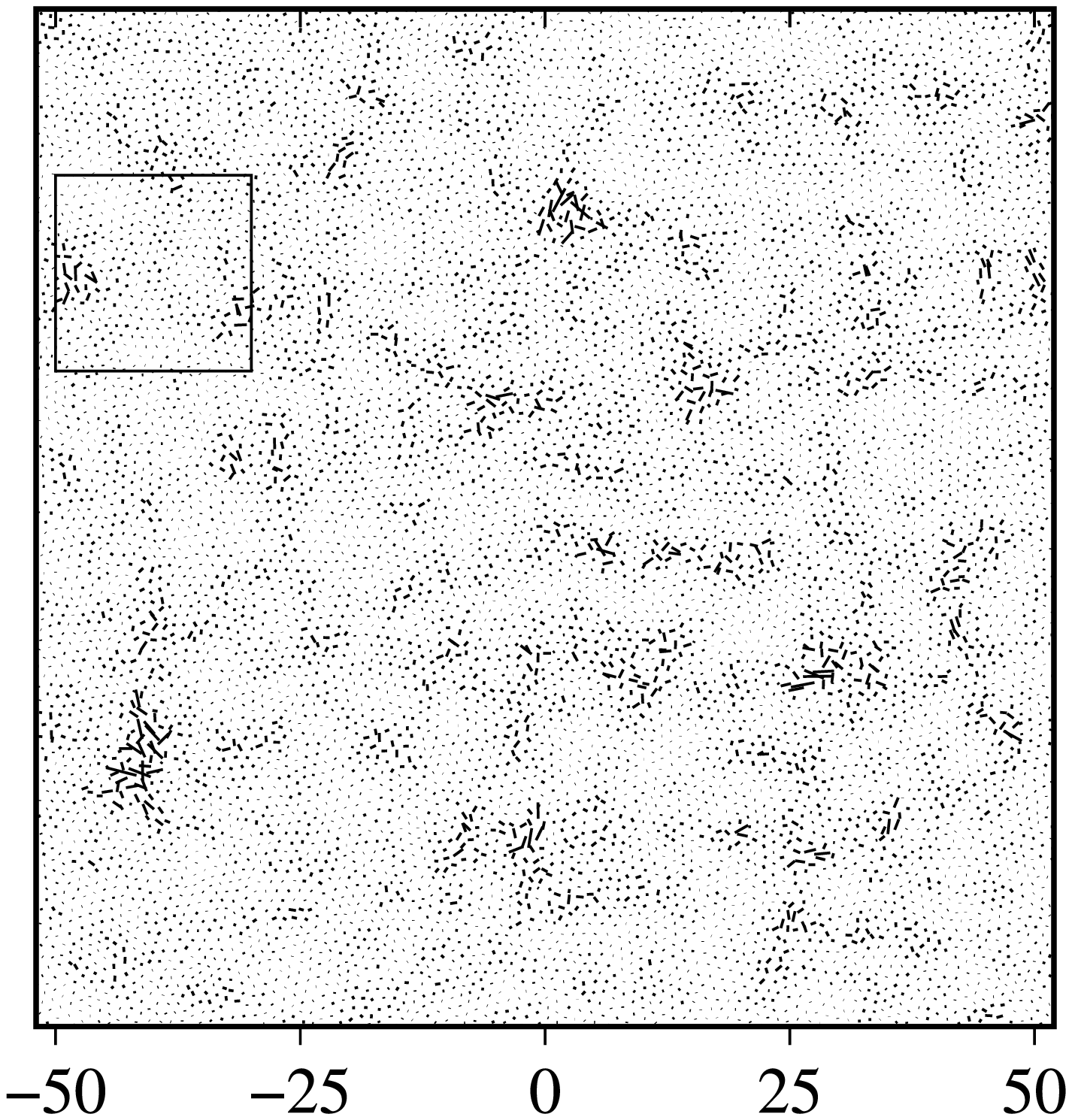}
\raisebox{0.435\hsize}{\bf d}
\includegraphics[width=0.475\hsize,clip]{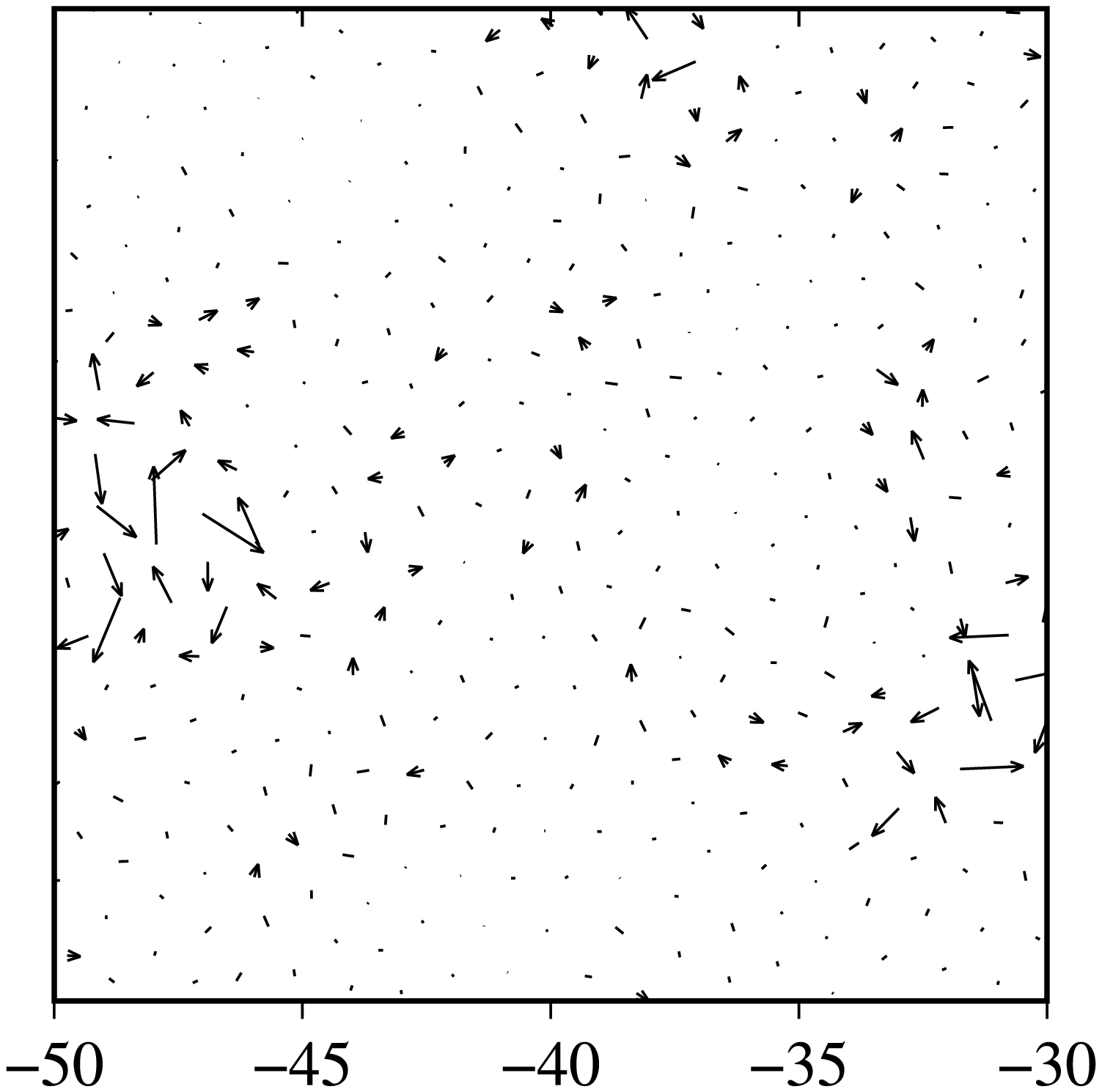}
\caption{{a. the non-affine displacements, $\delta\vect{u}_i$; c. the
    displacement fluctuations, $\vect{u}'_i$ (with CG width \mbox{$w=1$}) in a
    $10000$ particles 2D polydisperse Lennard-Jones glass under simple shear
    \mbox{$\gamma=5\cdot 10^{-7}$} (all vectors are magnified by $5\cdot
    10^{5}$); b,d. a corresponding zoom on a region of size $20\times 20$
    particles.}
  \label{fig:cgfields}}
\end{figure}

{Figs.~\ref{fig:cgfields}a,b show the non-affine particle
  displacements, $\delta\vect{u}_i$, in one configuration under simple shear.
  The vortex-like structures are qualitatively similar to those observed in
  simulations and experiments on amorphous materials
  (e.g.,~\cite{Langer97,Radjai02,Kolb04,Leonforte04,Leonforte05,Leonforte06}).
  The corresponding displacement fluctuations, $\vect{u}'_i$, calculated using
  a Gaussian CG function with \mbox{$w=1$}, are shown in
  Figs.~\ref{fig:cgfields}c,d (the choice of $w$ is discussed below). The
  fluctuations exhibit quite a different behavior from the non-affine field:
  they are more localized and considerably less correlated (as discussed
  below)}. The degree of localization can be quantified by the participation
ratio, defined as \mbox{$p\equiv \left( \sum_i{|\vect{u}'_i|^2} \right)^2 /
  \left[ N \sum_j{(|\vect{u}'_j|^2)^2} \right]$}.  Averaging over an ensemble
of 3 independent deformation {modes} for each of 9 different
realizations, we obtain (with \mbox{$w=1$}) \mbox{$p_{u'}=0.08\pm 0.02$}, while
for the non-affine field \mbox{$p_{\delta u}=0.28\pm 0.06$} (the error is the
standard deviation in the ensemble); {the same ensemble is used for
  Figs.~\ref{fig:correlation}, \ref{fig:corrlength_participation} and
  \ref{fig:component_distributions} below}.  The mean magnitude of the
fluctuations is also smaller, by a factor of about 4 ($4.4\cdot10^{-7}$ vs.\
$1.7\cdot10^{-6}$).

\begin{figure}[h!]
\raisebox{0.36\hsize}{\bf a}
    \includegraphics[height=0.39\hsize,clip]{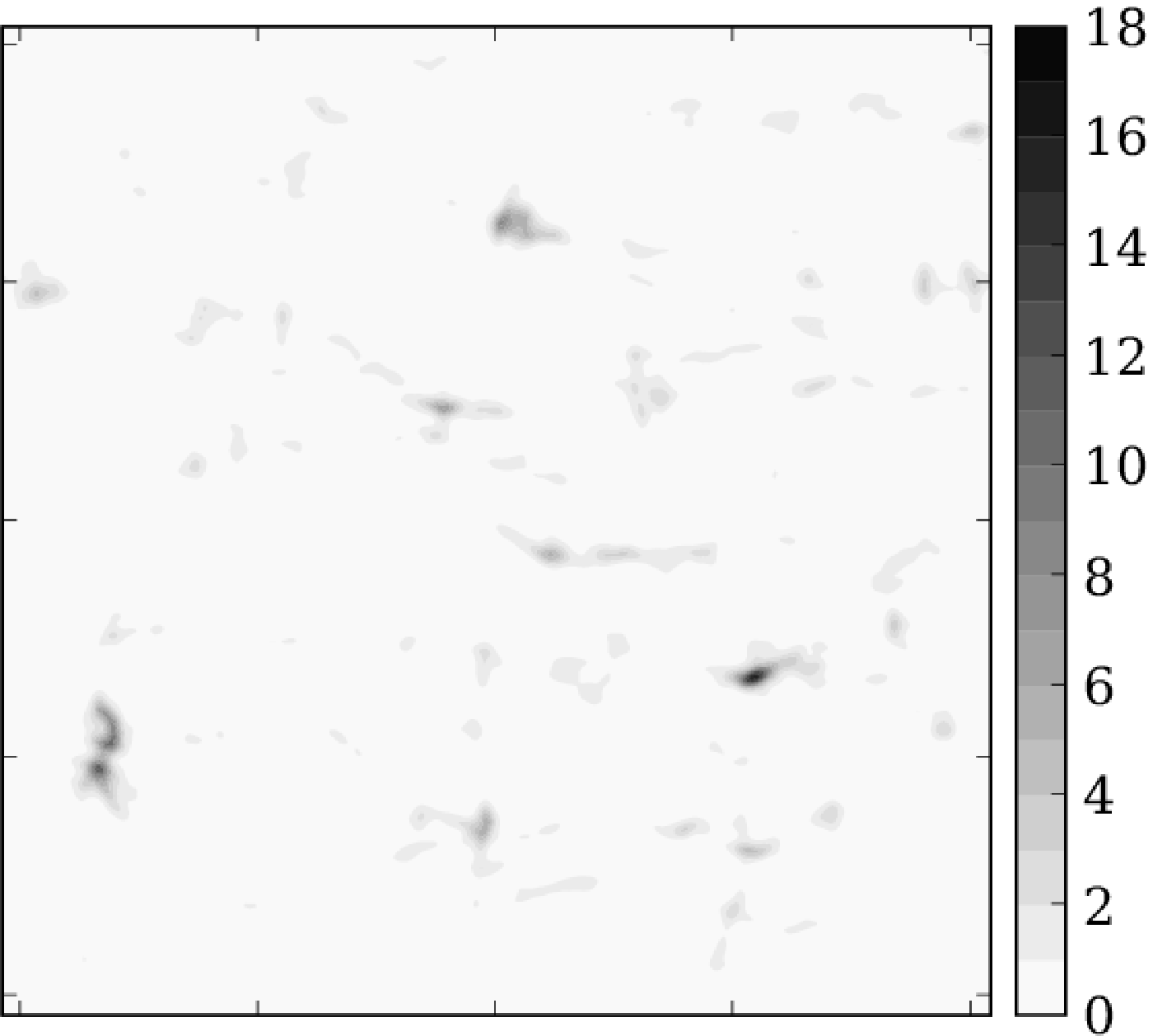}
\raisebox{0.36\hsize}{\bf b}\hspace{0.003\hsize}
    \includegraphics[height=0.39\hsize,clip]{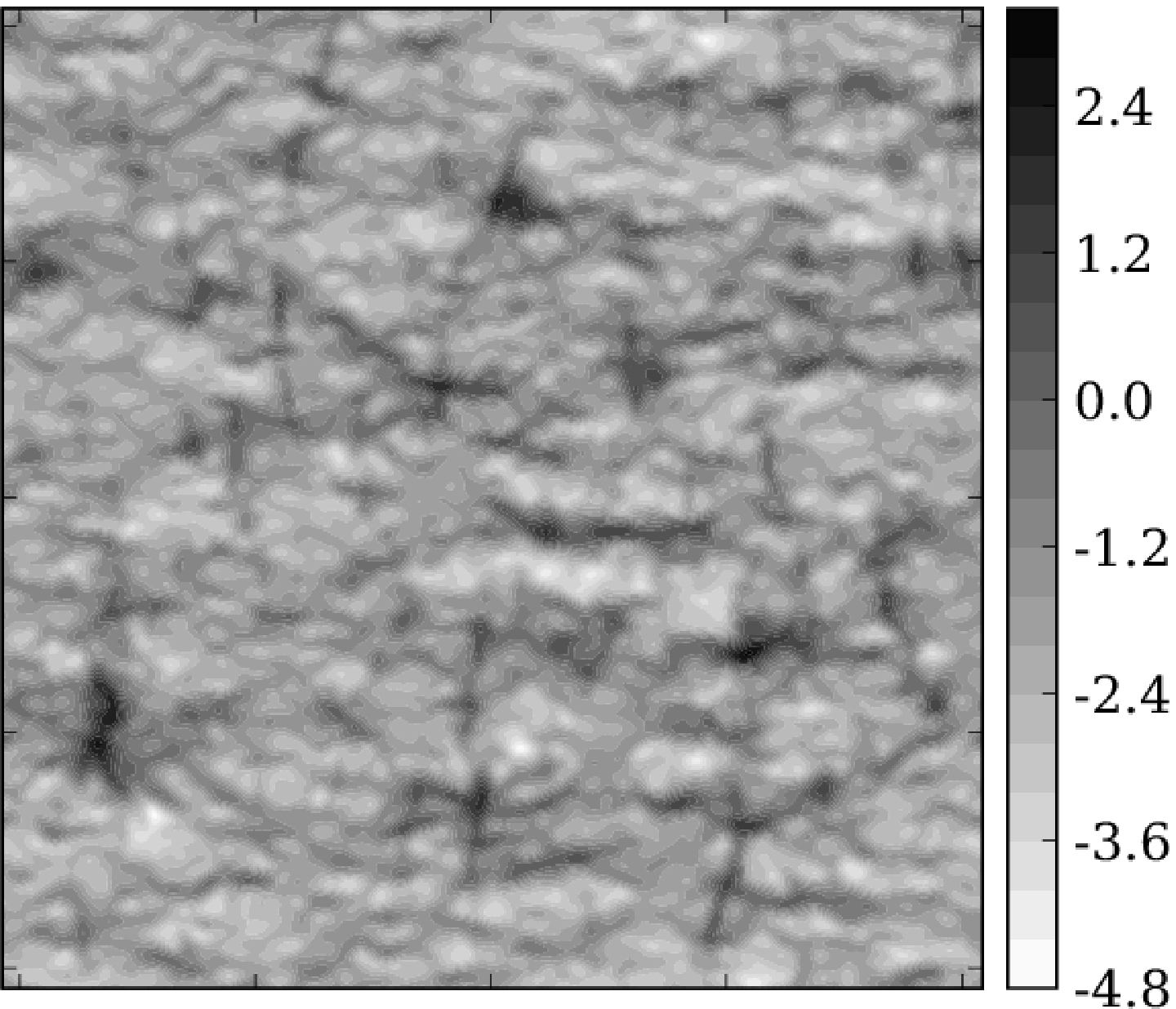}
\smallskip\\
\raisebox{0.36\hsize}{\bf c}
    \includegraphics[height=0.4\hsize,clip]{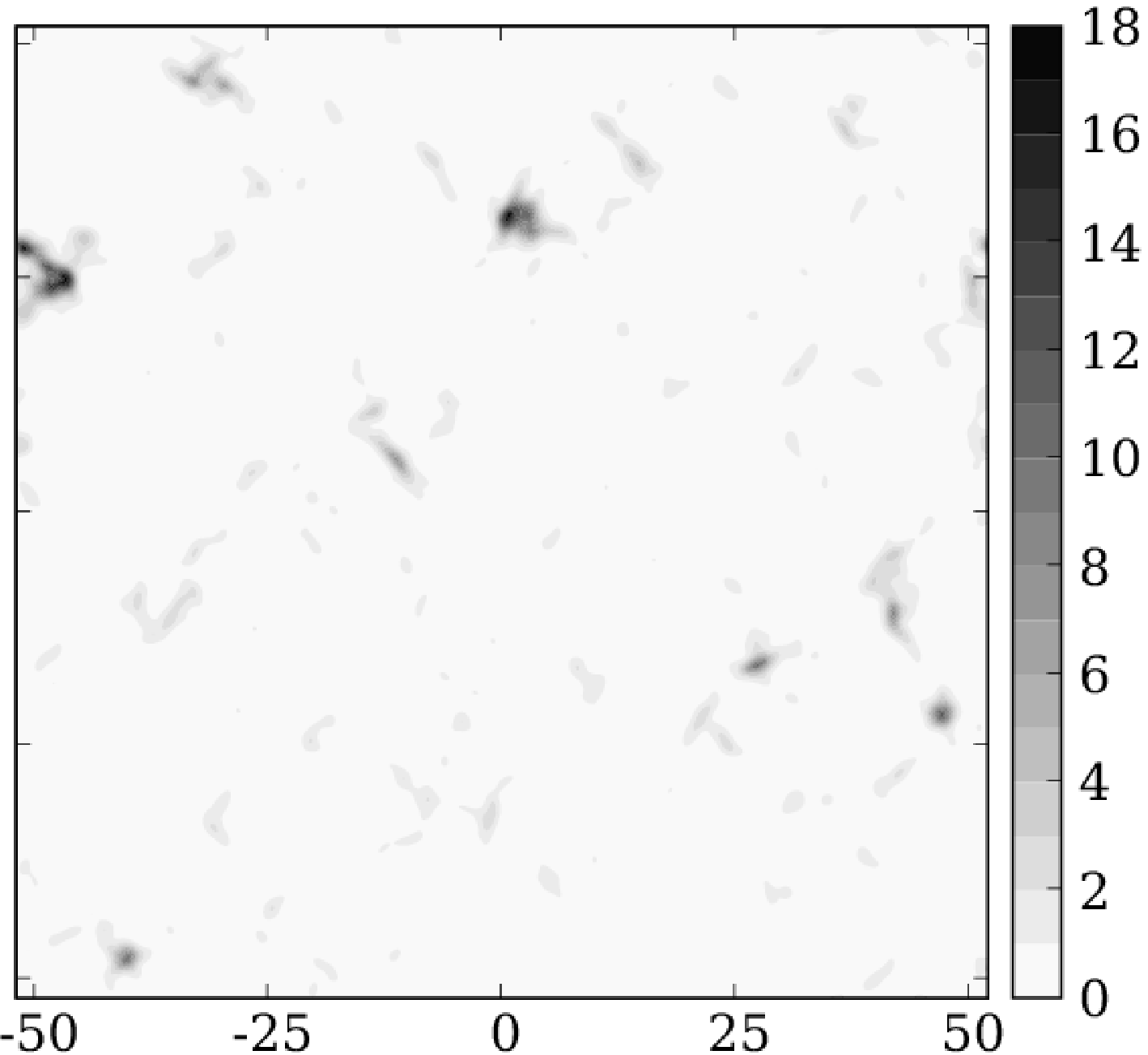}
\raisebox{0.36\hsize}{\bf d}
    \includegraphics[height=0.4\hsize,clip]{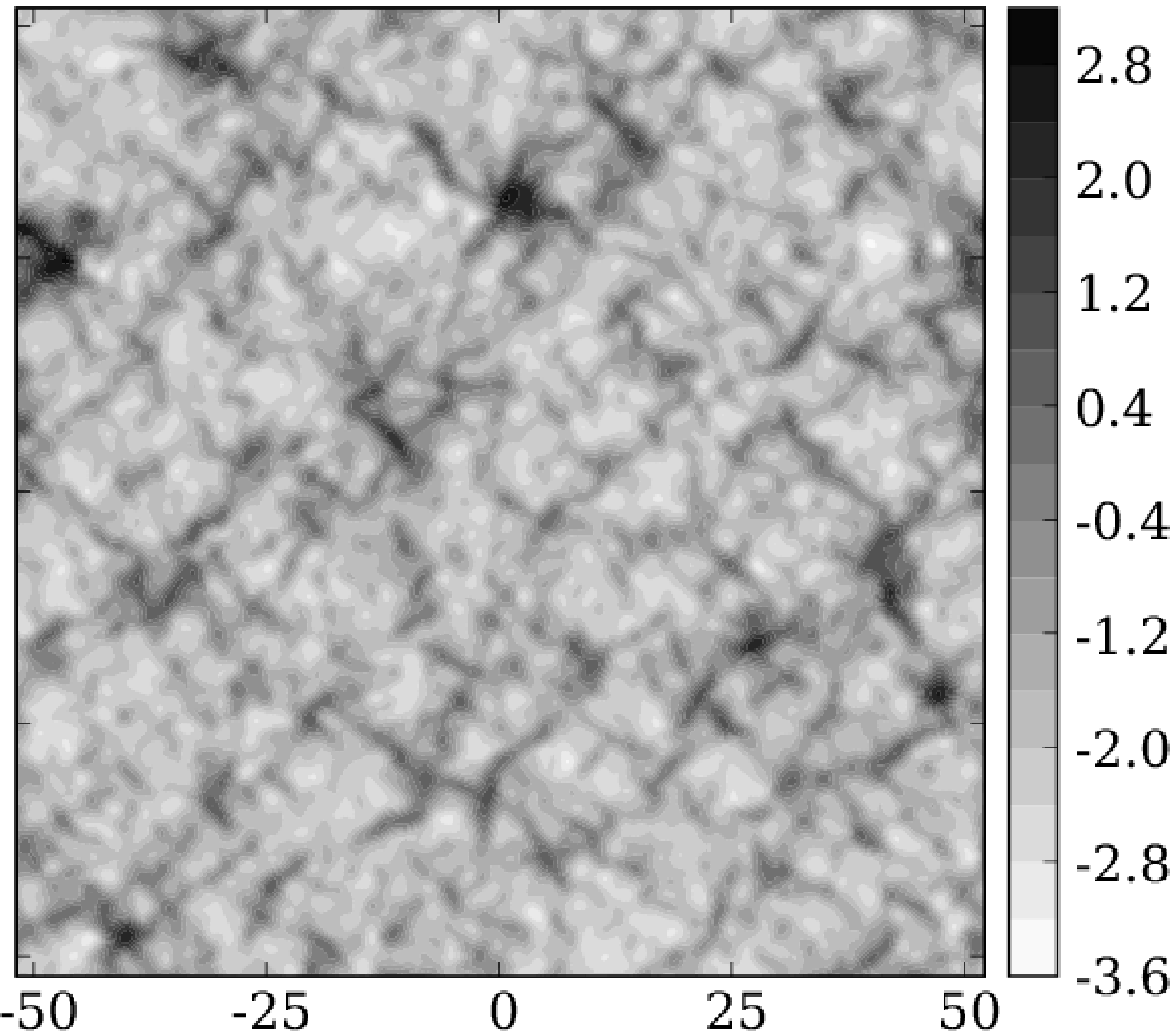}
\caption{Contour plots of (a,c) the noise field, $\eta(\vect{r})$ (with 
  \mbox{$w=1$}), in the system shown in Fig.~\ref{fig:cgfields}, scaled by
  $2.5\cdot 10^{11}$ and (b,d) its logarithm, for two different applied
  deformations: simple shear (a,b) and uniaxial stretching
  parallel to the $x$-axis (c,d). The noise field is calculated on a grid whose
  spacing is $w/4=1/4$. 
  \label{fig:noise}}
\end{figure}

\begin{figure}[h!]
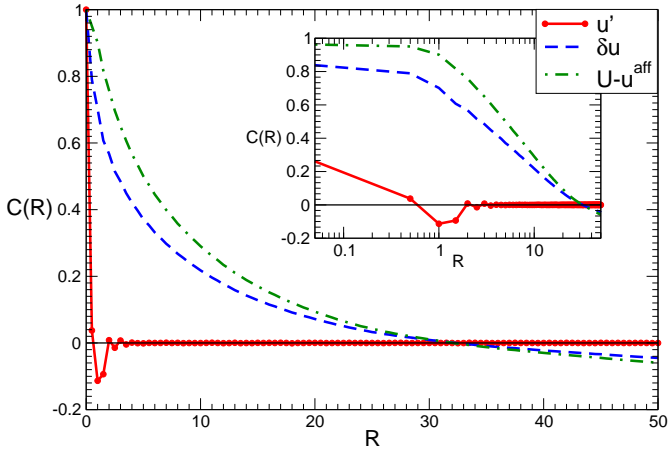

\begin{center}
  \onefigure[width=\hsize,clip]{pddam_v6_fig9.eps}
  \end{center}
  \caption{{Main:} ensemble averaged correlation of the scalar product of the
    non-affine displacements, $\delta\vect{u}_i$, the CG displacements at the
    particle positions, $\vect{U}(\vect{r}^0_i)$ (with \mbox{$w=1$}, the affine
    part subtracted) and the displacement fluctuations, $\vect{u}'_i$; inset:
    same graph with logarithmic horizontal scale.\label{fig:correlation}}
\end{figure}

The ``noise'' field, $\eta(\vect{r})$, presented in Fig.~\ref{fig:noise}, shows
very clearly the localized nature of the fluctuations. It exhibits rather sharp
peaks, whose positions depend both on the configuration and on the applied
deformation (e.g., shear or stretching). Preliminary
results~\cite{GoldenbergTanguyBarratIP} indicate that these peaks are
correlated with the positions of localized plastic rearrangements observed in
plastic deformation~\cite{Tanguy06}. Plotting the logarithm of the noise
(Fig{s}.~\ref{fig:noise}b,d) reduces the contrast and reveals
{anisotropic} patterns oriented predominantly near the principal shear
directions (parallel to the axes for simple shear, Fig.~\ref{fig:noise}b, and
at $45^{\circ}$ to the axes for uniaxial stretching, Fig.~\ref{fig:noise}d).
These correspond to the typical directions of the localized shear bands
observed in the plastic regime~\cite{Tanguy06}, which provides further support
to the notion that fluctuations and inhomogeneities in the elastic regime are
related to subsequent plastic failure and flow.

Another striking difference between the displacement fluctuations and the
non-affine field is observed in their correlation. Fig.~\ref{fig:correlation}
shows the scalar product correlation function, \mbox{$C(R)=\sum_{ij} \vect{u}_i
  \cdot \vect{u}_j \delta(R-|\vect{r}_i-\vect{r}_j|) / \sum_{ij}
  \delta(R-|\vect{r}_i-\vect{r}_j|)\,$}, normalized by its value at $R=0$
(i.e., the mean squared vector magnitude) for the three fields.  The
correlation of the non-affine field appears to be logarithmic on intermediate
scales. This behavior is captured quite well by the CG displacement with $w=1$.
The characteristic scale for the correlation of the non-affine field (for this
system size, $104\times 104$) is about $30$, similar to the diameter of the
vortices (Fig.~\ref{fig:cgfields}), while for the displacement fluctuations it
is {on the order of} one diameter.  {This justifies the assumption, made
  implicitly in~\cite{Goldhirsch02}, of the absence of long range correlations
  in the fluctuations, required for demonstrating that {\em local} linear
  elasticity applies on scales sufficiently large compared to their correlation
  length.}

\begin{figure}[h!]
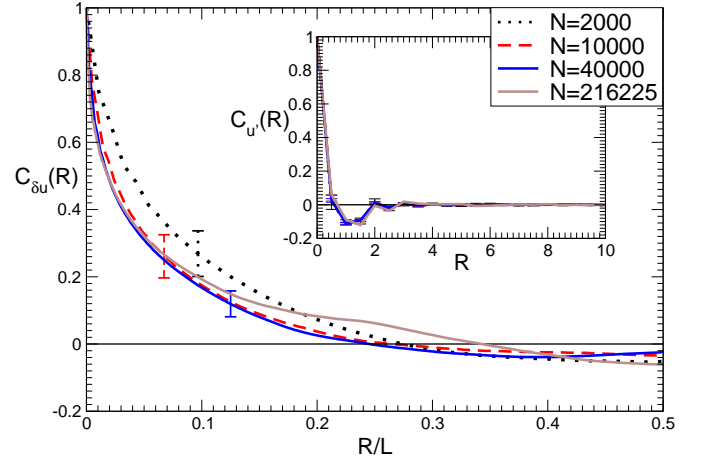

\begin{center}
  \onefigure[width=\hsize,clip]{pddam_v6_fig10.eps}
  \end{center}
  \caption{Main: ensemble averaged correlation of the scalar product of the
    non-affine displacements, $\delta\vect{u}_i$, for {systems of
      different sizes, subject to uniaxial stretching parallel to the $x$
      axis,} with the distance $R$ scaled by the length of the system, $L$;
    inset: the correlation of the displacement fluctuations, $\vect{u}'_i$
    (with \mbox{$w=1$}), in the same systems, distance
    unscaled.\label{fig:sizedep}}
\end{figure}

\begin{figure}[h!]
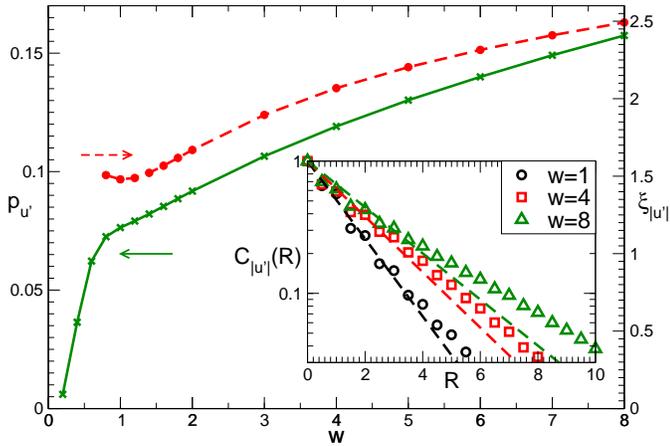

  \begin{center}
    \onefigure[width=\hsize,clip]{pddam_v6_fig11.eps}
  \end{center}
  \caption{{Main:} correlation length, $\xi_{{|u'|}}$, and
    participation ratio, $p_{u'}$, of the displacement fluctuations,
    $\vect{u}'_i$, vs.\ the CG width, $w${; inset: ensemble averaged
      correlation of the magnitude of the displacement fluctuations,
      $|\vect{u}'_i|$, for three values of $w$ (symbols); dashed lines
      indicate the exponential fits used to calculate
      $\xi_{{|u'|}}$}.\label{fig:corrlength_participation}}
\end{figure}

\begin{figure}[h!]
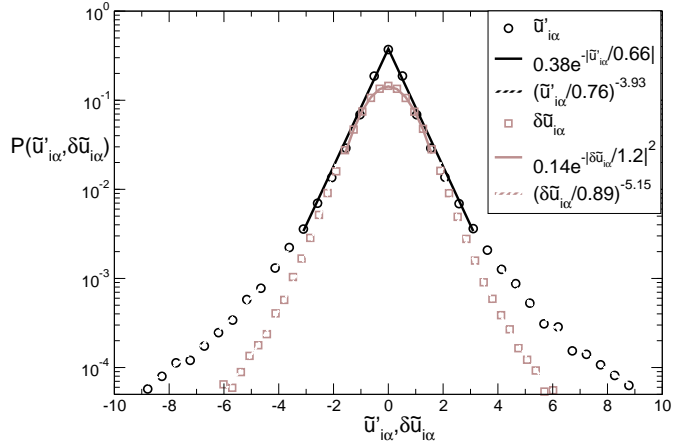

  \begin{center}
    \onefigure[width=\hsize,clip]{pddam_v6_fig12.eps}
  \end{center}
  \caption{Ensemble averaged distributions of both components of the non-affine
    displacement, {$\delta \tilde{u}_{i\alpha}\equiv\delta\vect{u}_{i\alpha} /
    \sigma_{\delta u}$, and of the displacement fluctuations,
    $\tilde{u}'_{i\alpha}\equiv\vect{u}'_{i\alpha}/\sigma_{u'}$ (with
    \mbox{$w=1$}), where \mbox{$\sigma_{\delta u}=1.31\cdot 10^{-6}$} and
    \mbox{$\sigma_{u'}=4.05\cdot 10^{-7}$} are the corresponding standard
    deviations.}\label{fig:component_distributions}}
\end{figure}

As mentioned, the correlation of the non-affine field has been observed to
depend on the system size, in a somewhat different system of bidisperse disks
with harmonic repulsion~\cite{Maloney06b}. We verified that this is indeed the
case in the systems we consider here: Fig.~\ref{fig:sizedep} presents the
scalar product correlation function of the non-affine field and of the
displacement fluctuations for four system sizes: $46.5\times 46.5$ ($N=2000$),
$104\times 104$ ($N=10000$), $208\times 208$ ($N=40000$) and $483.5 \times
483.5$ ($N=216225$), a considerably larger range of sizes than considered
in~\cite{Maloney06b}. The results were obtained by averaging over $20$, $8$ and
$8$ different realizations respectively for the first three sizes (only one
configuration with $N=216225$ was used), {for one deformation mode:}
uniaxial stretching parallel to the $x$ axis; {representative error
  bars indicate the standard deviation in the ensemble}. The correlation of the
non-affine field exhibits a rather good collapse (at least for the larger
systems considered) when the distance is scaled by the length of the system,
$L$ (the deviation observed for {$N=216225$} may be due to reduced
statistics, or to {two weak ``plastic events'' which occurred in the
  system, indicating a possible deviation from a linear response}).  On the
other hand, as shown in the inset of Fig.~\ref{fig:sizedep}, the correlation of
the displacement fluctuations does not depend on the system size, i.e., unlike
the non-affine field, it provides a {\em local} characterization of the
disorder in the particle displacements.

The CG displacement and its fluctuations obviously depend on the chosen CG
scale (resolution), $w$. For $w \to 0$, $\vect{U}(\vect{r}^0_i)=u_i$ so that
$\vect{u}'_i=0$. For $w \to \infty$, $\vect{U}(\vect{r}^0_i)=0$, since the
center of mass of the system is fixed, hence $\vect{u}'_i=\vect{u}_i$. In order
to obtain the optimal continuum description we would like to choose $w$ to be
as small as possible. In lattices with a complex unit cell, the deformation can
only be expected to be locally homogeneous, or affine, on scales larger than
the size of the unit cell. In our case there is no periodicity in the
structure; hence, by analogy, the minimum scale on which we can expect to
define a meaningful CG displacement is on the order of the interparticle
separation, $w\simeq\left<\sigma\right>=1$. {The oscillatory nature of
  the correlation of the scalar product of $\vect{u}'_i$
  (Figs.~\ref{fig:correlation},\ref{fig:sizedep}) makes it difficult to define
  a correlation length; we therefore use the correlation of the {\em
    magnitudes}, $|\vect{u}'_i|$, which is fit rather well by
  $C_{|u'|}(R)=\exp(-R/\xi_{|u'|})$, at least for sufficiently small $w$; see the
  inset of Fig.~\ref{fig:corrlength_participation}}.  In
Fig~\ref{fig:corrlength_participation}, we present the dependence on $w$ of the
participation ratio, $p_{u'}$, and of the correlation length $\xi_{|u'|}$.  As
expected, the results are scale dependent: the goal is to characterize the {\em
  inhomogeneity} of the displacement field in particular configurations; scale
independence is expected only for locally homogeneous fields, or for ensemble
averages when such averages are homogeneous~\cite{Goldenberg06b}.  The
participation ratio increases sharply with $w$ up to $w\simeq 1$, above which
it increases more slowly, eventually reaching (for large $w$; not shown) values
similar to those of the non-affine field, as expected. {The
  correlation length, $\xi_{|u'|}$, shows a small ``plateau'' around $w\simeq
  1$, at a value of about $1.5$ mean diameters. This supports our choice of
  \mbox{$w=1$} for separating the smooth, continuous part of the displacement
  from the ``noise''. As expected, $\xi_{|u'|}$ increases with $w$ (due to the
  long range correlation of the CG displacement, hence the reduced quality of
  the exponential fit), but remains much smaller than the typical length
  observed for the non-affine field
  (Figs.~\ref{fig:correlation},\ref{fig:sizedep}).}

It is interesting to compare the ensemble averaged distributions of the
components of the displacement fluctuations (with \mbox{$w=1$}) and of the
non-affine field, shown in Fig.~\ref{fig:component_distributions} (averaged
over {the $x$ and $y$ components, whose distributions are the same
  within statistical error, i.e., they are isotropic}).  The two distributions
are qualitatively different.  In particular, the center of the distribution is
exponential for the fluctuations (the fit shown in
Fig.~\ref{fig:component_distributions} is for
${\tilde{u}'_{i\alpha}}<3.5$), but Gaussian (the fit shown is for ${\delta
\tilde{u}_{i\alpha}}<1.6$) for the non-affine field (see
also~\cite{Radjai02}).  Both distributions exhibit power law tails, with
different exponents. For $w\gtrsim 1$, we find (not shown) that the
distribution of the fluctuations crosses over to a Gaussian at the center,
presumably due to the contribution of the continuous part of the displacement
(which is dominant in the distribution of the non-affine field).  This provides
further support for the choice of \mbox{$w=1$}.  Note that the crossover to a
Gaussian distribution is not due to the Gaussian CG function used: a Heaviside
CG function yields similar results.

In~\cite{Tanguy02} {(Fig.~10)}, a length $\xi\simeq 30\left< \sigma \right>$,
independent of the system size, was found to characterize the crossover of the
vibrational modes to their continuum limit, in systems similar to the ones
studied here.  A similar crossover length seems to apply for a uniform applied
strain~\cite{Goldhirsch02,Tanguy02}. It is not clear how, and whether, this
length is related to the disorder in the displacement field. The correlation
{lengths} of the fluctuations (Fig.~\ref{fig:sizedep}), as shown above, or of
the noise field {(about $2.1$ with \mbox{$w=1$}), are much smaller than $\xi$},
and are system size independent; {however,} the correlation {of these fields}
does not clearly reveal an {\em additional} length.  The {\em distribution} of
the fluctuations (or noise) is also essentially size independent:
Fig.~\ref{fig:noise_distributions} presents the distribution of the noise field
for systems of different sizes (averaged over an ensemble of configurations,
except for the larger system; the deviations observed for the latter are
probably due {to the ``plastic events'' mentioned above, around which the noise
  is larger}).  {We verified the size independence of the distributions of the
  fluctuations and of the non-affine field. The enhancement of small noise for
  the sheared systems is also observed in the components of $\vect{u}'_{i}$,
  for which the probability is larger by about $10\%$ for
  $\tilde{u}'_{i\alpha}\lesssim 0.2$; this effect is too small to be seen in
  Fig.~\ref{fig:component_distributions}, which justifies the averaging over
  deformation modes used in that figure.  The patterns observed in
  Fig{s}.~\ref{fig:noise}b,d probably reflect a more significant difference in the
  {\em correlation} for different deformation modes rather than this small
  difference in the distribution. {Defining a density of ``defects'',
    i.e., regions of large noise (the peaks visible in Figs.~\ref{fig:noise}a,c)
    would enable the definition of a corresponding length. However, the
    objective identification of the defects is difficult, since the power law
    decay of the noise distribution precludes the definition of a natural
    cutoff for the noise. We have therefore not been able to clearly identify
    the mesoscopic length scale $\xi$ in the fluctuations (or the non-affine
    field). This remains an important issue for further study.}
\begin{figure}[h!]
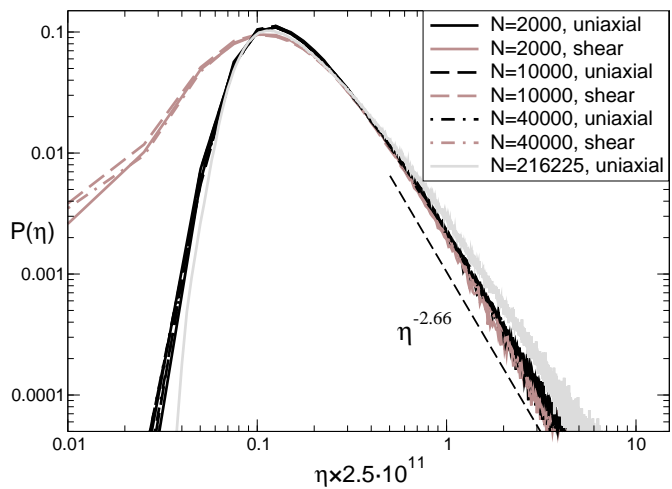

  \begin{center}
    \onefigure[width=\hsize,clip]{pddam_v6_fig13.eps}
  \end{center}
  \caption{Ensemble averaged distributions of the noise field, $\eta(\vect{r})$
    (with \mbox{$w=1$}), scaled by $2.5\cdot 10^{11}$, in systems of different
    size, subject to either uniaxial stretching or simple
    shear.\label{fig:noise_distributions}}
\end{figure}

\section{Conclusion}
{Our results suggest that the main features of the non-affine field
  may be described by an inhomogeneous continuum model, e.g., local linear
  elasticity (as in~\cite{DiDonna05}) which should apply on sufficiently large
  scales, on the order of ten
  diameters~\cite{Goldhirsch02,Tanguy02,Leonforte04,Leonforte05}; however,
  corrections to local elasticity are expected to be required on smaller
  scales. Both the non-affine field and the ``intrinsic'' microscopic
  fluctuations exhibit wide tailed distributions; however, the latter exhibit
  much shorter correlations. While the fluctuations cannot be described by
  classical continuum models, their short correlation should facilitate a
  statistical description, possibly in terms of local ``defects''.} We propose
a related ``noise'' field which may be used for extending {such
  models}, e.g., to describe the part of the elastic energy which is not
captured by the CG displacement~\cite{Goldhirsch02,Serero07}.  The localized
nature of the ``noise'' field, as well as the patterns it exhibits, suggest a
relation to the localized plastic rearrangements observed in plastic
deformation~\cite{Tanguy06}. This relation is confirmed by preliminary
results~\cite{GoldenbergTanguyBarratIP}, and is currently being studied in
further detail. Such a relation may provide a (currently lacking) microscopic
basis for phenomenological models for plastic flow involving localized
rearrangements of zones which interact via the elastic deformation of the
material, such as the Shear Transformation Zone (STZ) model~\cite{Falk98} and
similar models (e.g.,~\cite{Picard05,Baret02}). Both inhomogeneities
(pertaining to the continuum description) and microscopic fluctuations may be
important in this context. The approach suggested here should be useful both
for analyzing simulations, as described in this Letter, and experimental data.

\begin{acknowledgments}
  We thank F.~L{\'e}onforte and M.~Tsamados for providing the reference
  configurations used in this study, and I.~Goldhirsch for useful discussions.
  C.~G.\ acknowledges financial support from the R{\'e}gion Rh{\^o}ne-Alpes,
  France and from a European Community FP6 Marie Curie Action
  (MEIF-CT2006-024970).
\end{acknowledgments}

\end{document}